\documentclass{IEEEcsmag}
\usepackage[utf8]{inputenc}
\usepackage[colorlinks,urlcolor=magenta,linkcolor=black,citecolor=black]{hyperref}
\usepackage{upmath}
\usepackage{xspace}
\usepackage{ifthen}
\usepackage[T1]{fontenc}
\usepackage{xcolor}
\usepackage{amsmath,amssymb}
\usepackage{svg}
\usepackage{booktabs}
\usepackage{multirow}
\usepackage{icomma}
\usepackage{fp}
\usepackage[autolanguage]{numprint}
\usepackage[scaled=0.85]{beramono}
\usepackage{tikz}
\usepackage{listings}
\usepackage{multicol}
\usepackage{amsmath}
\usepackage{balance}

\jvol{XX}
\jnum{XX}
\paper{XX}
\jmonth{XX/XX}
\jname{}
\pubyear{2022}

\definecolor{pblue}{rgb}{0.13,0.13,1}
\definecolor{pgreen}{rgb}{0,0.5,0}
\definecolor{pred}{rgb}{0.9,0,0}
\definecolor{pgrey}{rgb}{0.46,0.45,0.48}
\usepackage{listings}
  \lstdefinelanguage{diff}{
    basicstyle=\ttfamily\small,
    morecomment=[f][\color{pgrey}]{@@},
    morecomment=[f][\color{pred}]{+\ },
    morecomment=[f][\color{pgreen}]{-\ },
  }

\makeatletter
\newenvironment{btHighlight}[1][]
{\begingroup\tikzset{bt@Highlight@par/.style={#1}}\begin{lrbox}{\@tempboxa}}
{\end{lrbox}\bt@HL@box[bt@Highlight@par]{\@tempboxa}\endgroup}
\newcommand\btHL[1][]{%
  \begin{btHighlight}[#1]\bgroup\aftergroup\bt@HL@endenv%
}
\def\bt@HL@endenv{%
  \end{btHighlight}%
  \egroup
}
\newcommand{\bt@HL@box}[2][]{%
  \tikz[#1]{%
    \pgfpathrectangle{\pgfpoint{1pt}{0pt}}{\pgfpoint{\wd #2}{\ht #2}}%
    \pgfusepath{use as bounding box}%
    \node[anchor=base west, fill=orange!30,outer sep=0pt,inner xsep=1pt, inner ysep=0pt, rounded corners=0pt, minimum height=\ht\strutbox+1pt,#1]{\raisebox{1pt}{\strut}\strut\usebox{#2}};
  }%
}
\lstdefinestyle{Java}{
    language={Java}, 
    moredelim=**[is][{\btHL[fill=red!17,thin]}]{`}{`},
    moredelim=**[is][{\btHL[fill=green!17,thin]}]{@}{@},
    moredelim=**[is][{\btHL[fill=yellow!17,thin]}]{~}{~},
}

\newboolean{showcomments}
\setboolean{showcomments}{true}
\ifthenelse{\boolean{showcomments}}
{
}

\newcommand{\aka}{a.k.a.\@\xspace}
\newcommand{\eg}{e.g.,\@\xspace}

\newcommand*\np[2][z]{
        \textcolor{black}{%
		\ifx z#1%
		$\numprint{#2}$%
		\else%
		$\numprint[#1]{#2}$%
		\fi\xspace
		}
}

\newcommand{\ShowAbsoluteNumber}[1]{%
\ifnum #1<10%
{\hspace*{0pt}#1}%
\else%
\ifnum #1<100%
{\hspace*{0pt}#1}%
\else%
\ifnum #1<1000%
{\hspace*{0pt}#1}%
\else%
{\numprint{#1}}%
\fi%
\fi%
\fi%
}

\newcommand{\ShowPercentage}[2]{%
\FPeval\percentage{round(#1/#2*100,0)}%
\FPeval\percentageOneDecimal{round(#1/#2*100,1)}%
\ifnum \percentage=0%
{\np[\%]{\FPprint{percentageOneDecimal}}}%
\else%
\ifnum \percentage<10%
{\np[\%]{\FPprint{percentageOneDecimal}}}%
\else%
{\np[\%]{\FPprint{percentageOneDecimal}}}%
\fi%
\fi%
\xspace
}
\newlength\BARSIZE  
\newcommand{\inlinechart}[2]{%
\FPeval{\BLACKBARSIZE}{#1/#2}\textcolor{black!80}{\rule{\BLACKBARSIZE\BARSIZE}{1.6ex}}%
\FPeval{\BLACKBARSIZE}{1 - (#1/#2)}\textcolor{black!10}{\rule{\BLACKBARSIZE\BARSIZE}{1.6ex}}%
}

\newcommand*\ChartSmall[3][v]{%
\ifx q#1%
    \np{#2}/\np{#3}(\ShowPercentage{#2}{#3})\else%
\ifx p#1%
    \np{#2}(\ShowPercentage{#2}{#3})\else%
\ifx c#1%
    \inlinechart{#2}{#3}%
\else%
    \np{#2}%
    \ifx r#1%
        /\np{#3}%
    \fi%
    \hspace*{0.5ex}(\ShowPercentage{#2}{#3}) %
    \inlinechart{#2}{#3}%
    \xspace
\fi\fi\fi%
}

\begin{document}

\title{The Multibillion Dollar Software Supply Chain of Ethereum}
\author{César Soto-Valero \href{https://orcid.org/0000-0003-0541-6411}{\includegraphics[scale=0.05]{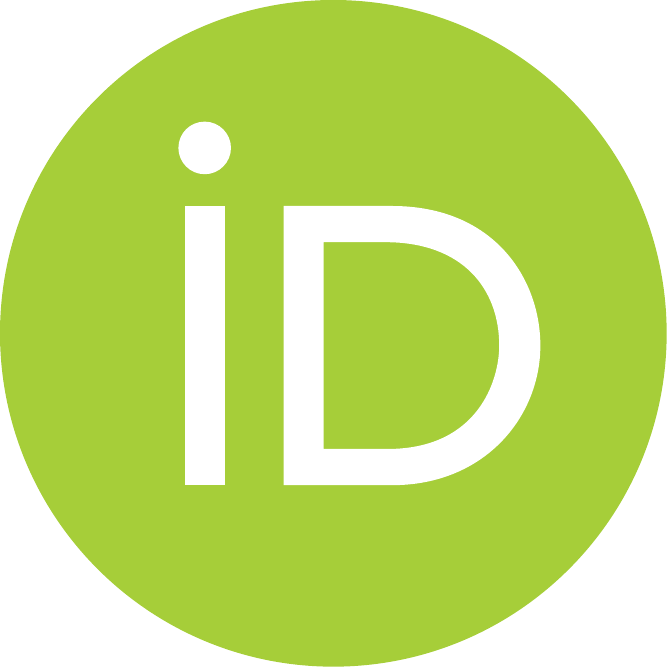}}}
\affil{KTH Royal Institute of Technology}
\author{Martin Monperrus \href{https://orcid.org/0000-0003-3505-3383}{\includegraphics[scale=0.05]{ORCID-iD_icon-vector.pdf}}}
\affil{KTH Royal Institute of Technology}
\author{Benoit Baudry \href{https://orcid.org/0000-0002-4015-4640}{\includegraphics[scale=0.05]{ORCID-iD_icon-vector.pdf}}}
\affil{KTH Royal Institute of Technology}

\begin{abstract}
The rise of blockchain technologies has triggered tremendous research interest, coding efforts, and monetary investments in the last decade.
Ethereum is the single largest programmable blockchain platform today. It features cryptocurrency trading, digital art, and decentralized finance through smart contracts.
So-called Ethereum nodes operate the blockchain, 
relying on a vast supply chain of third-party software dependencies maintained by diverse organizations.
These software suppliers have a direct impact on the reliability and the security of Ethereum.
In this article, we perform an analysis of the software supply chain of Java Ethereum nodes and distill the challenges of maintaining and securing this blockchain technology.
\end{abstract}

\maketitle

\chapterinitial{Ethereum} is a spearhead of the blockchain paradigm, with its smart contract infrastructure supporting a vibrant decentralized finance ecosystem~\cite{dannen2017introducing} and a blooming art scene \cite{monroe2022art}.
Since the release of Bitcoin in 2008~\cite{Nakamoto2008}, the adoption of blockchain-based solutions has grown significantly, mainly driven by the promise of secure, reliable, and decentralized monetary and financial transactions~\cite{wsjdefi}.
There are several public blockchains running today (\eg Bitcoin, Ethereum, Litecoin, and NEO), each one of them serving a particular purpose and solving specific problems. In this article, we focus on the single case of Ethereum, as it is the largest blockchain platform by most notable metrics.

Ethereum is a feature-rich platform, considered by some as the avant-garde of  blockchain technologies~\cite{Wood2014}.
It has its own cryptocurrency (Ether), its own consensus protocol, and its own smart contract platform.
Ethereum digital assets and contracts are executed in a distributed manner in nodes that support the Ethereum Virtual Machine (EVM) execution model. 
Due to this functionality, the Ethereum platform is often compared to a globally distributed supercomputer.
In January 2022, the Ethereum blockchain held over hundreds of billion U.S. dollars in digital assets, and an average of \np{250} new smart contracts are deployed and verified on Etherscan every day.\footnote{\url{https://etherscan.io/chart/verified-contracts}}

The research community has contributed to the creation and evolution of blockchain technologies.
The recent work focuses on three aspects \cite{Huang2021}: theoretical foundations, scalability, and engineering of smart contracts.
However, one key aspect of the blockchain has been completely overlooked: its software supply chain.

The software supply chain of an ecosystem is the set of all software libraries, tools and third-party modules that compose it~\cite{Lamb2021}.
In the context of Ethereum, the software supply chain is first and foremost formed of the different open-source implementations of Ethereum nodes, in Go, Rust, Java and other languages~\cite{Pooja2020}. 
The node implementations themselves depend on hundreds of components.
Overall, the software supply chain of Ethereum is composed of libraries and tools, \aka dependencies, to develop, deploy and run Ethereum nodes. 

Recent studies have shown that a large network of software dependencies such as in Ethereum nodes can turn into an application's Achilles heel~\cite{gkortzis2021software}.
On the one hand, malicious actors may infect a target application from within a reused component~\cite{Ohm2020}. 
On the other hand, entire software systems may crash because of a bug somewhere deep in the reuse chain~\cite{Massacci2021TechnicalLI}.
The major stakeholders in the Ethereum ecosystem want it to be resistant to attackers and robust with respect to bugs~\cite{Aumasson2021}. 
Consequently, understanding and hardening its software supply chain has become of utmost importance.

In this article, we deep dive into the software supply chain of the two main Ethereum nodes written in Java, namely Besu and Teku.
Our focus on Java is motivated by the strong presence of Java in banks and financial institutions, an essential target group of Ethereum, as well as by the availability of advanced tools for analyzing and hardening the supply chain in Java~\cite{Massacci2021TechnicalLI}.

We analyze the software supply chains of two Java Ethereum nodes, looking at their dependencies and their evolution over time.
This provides the community with the first ever description of a mission critical software supply chain for blockchain. 
Next,  we provide actionable results and show that we can harden a  complex software supply chain with relevant tools.
Our results reveal a number of key insights and technical challenges for both researchers in software supply chains and well as for developers and stakeholders of the Ethereum community.

\begin{figure*}[!t]
\centering
\includegraphics[width=16cm]{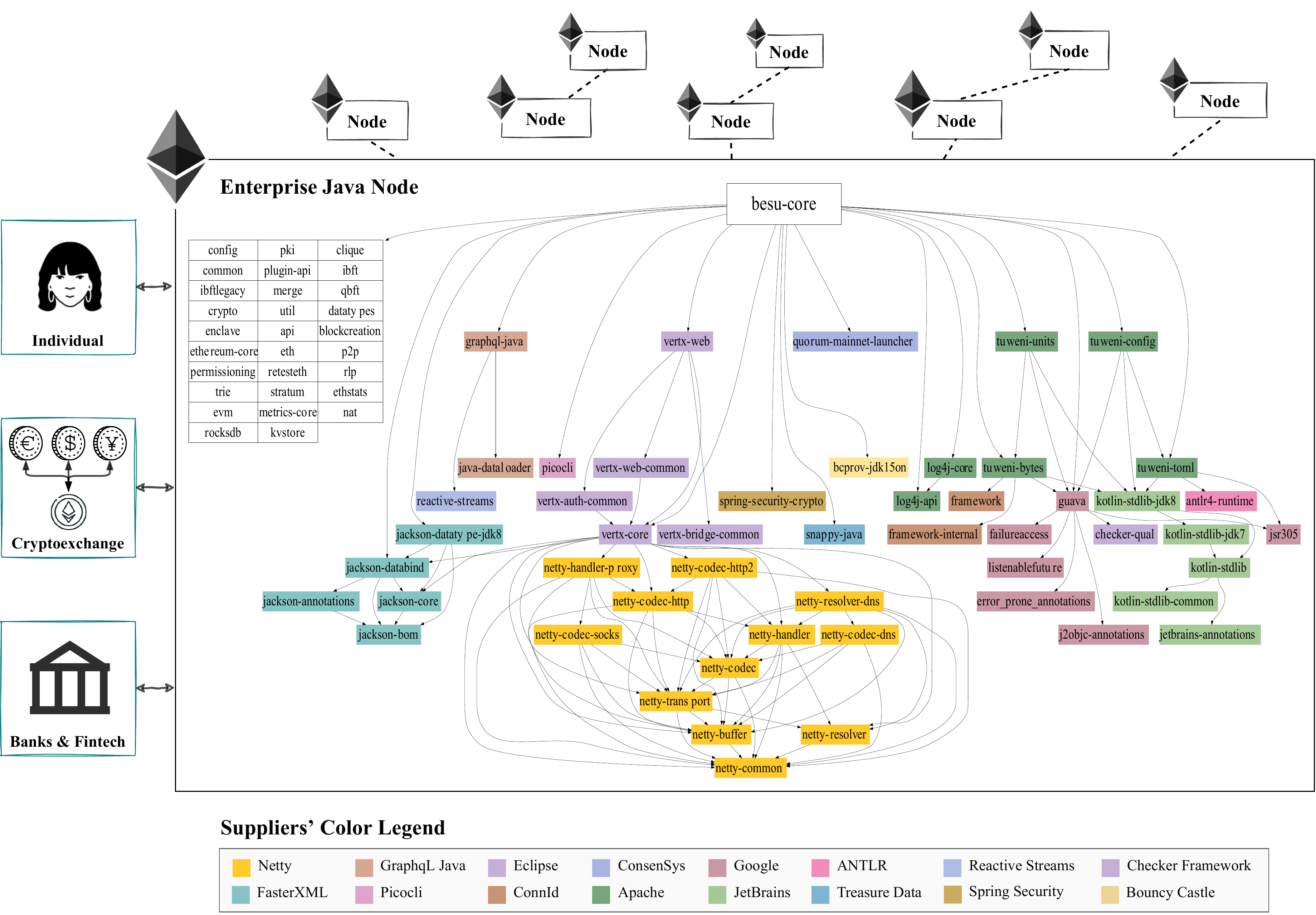}
\caption{Excerpt of the software supply chain of one single module in the enterprise Java Ethereum node Besu v21.10.6: the \texttt{besu-core} module depends on 29 other Besu modules, as well as on \np{51} third-party dependencies provided by \np{16} different supplier organizations; the third-party dependencies are colored according to the name of the supplier.}
\label{fig:besu_node}
\end{figure*}

\section{SUPPLY CHAIN OF JAVA ETHEREUM NODES}

\subsection{Overview}

Ethereum is a public distributed system of nodes supporting a ledger. 
As a distributed system, Ethereum nodes agree on a consensus protocol to verify the validity of transactions. 
The protocol for  Ethereum v1.0 (\textsc{Eth1}) is based on proof of work (PoW), and for Ethereum v2.0 (\textsc{Eth2}), it is based on proof of stake (PoS).
Ethereum nodes communicate peer-to-peer without a central organizing institution. 
They run smart contracts and receive transactions from client applications.  
This is what is depicted in the outer parts of \autoref{fig:besu_node}.

The top part of \autoref{fig:besu_node} illustrates the network of Ethereum nodes. 
The left part of the figure illustrates the three main categories of clients of the Ethereum blockchain: an individual user, a cryptoexchange marketplace, and a bank.
The individual user is, for example, an artist who relies on the blockchain to distribute her artwork \cite{jawn.eth}. 
The cryptoexchange marketplace provides deposits and withdrawals of the Ether cryptocurrency.
The bank uses Ethereum to accelerate payments across borders, opening up the possibility to help underbanked populations \cite{schuetz2020blockchain}.
The central part of \autoref{fig:besu_node} focuses on one single Ethereum node, and provides a deep dive into its software supply chain.

From now on, we assume this node runs the Java implementation of Ethereum called Besu.
\autoref{fig:besu_node} shows the graph of dependencies for the core module of Besu, which is only one of the 41 modules of Besu, focusing on its direct dependencies. 
Such a large number of software dependencies is a potential source of vulnerabilities and supply chain attacks~\cite{Pashchenko2018}.
In the context of Ethereum, this means that the software dependencies of Besu represent a potential source of risk for the financial  system and art market built on top of it. In the rest of this article, we study potential counter-measures.

\begin{table*}[htb]
\centering
\caption{Descriptive statistics of the software supply chain of the two major enterprise Java Ethereum nodes: Besu v21.10.6 (\href{https://github.com/hyperledger/besu/tree/ef7984b50529d07a74b3d9c6f73c51dfc8e84277}{ef7984b}) and Teku v22.1.0 (\href{https://github.com/ConsenSys/teku/tree/5b85ef197f5468c32b4aeb869ebe74201b9875bf}{5b85ef1}). }
\begin{tabular}{@{}lccc@{}}
\toprule
                                 & \textsc{Besu (Eth1)}  & \textsc{Teku (Eth2)}  \\ \midrule
\textsc{Lines of Java Code} &  \np{268356}            & \np{209860}        \\ 
\textsc{Commits} & \np{3125}            & \np{3142}              \\
\textsc{Contributors} & \np{115}            & \np{65}         \\ 
\midrule
\textsc{Unique Internal Dependencies}     & \np{41}             & \np{57}          \\ 
\textsc{Unique Third-party Dependencies} & \np{355}            & \np{293}       \\ 
\textsc{Unique Suppliers } & \np{165}    & \np{146}    \\
\midrule
\textsc{Unique Third-party Dependencies Introduced Since January 2021} & \np{127}  & \np{79}  \\ 
\textsc{Unique Third-party Suppliers Introduced Since January 2021} & \np{49}  & \np{22}  \\
\textsc{Unique Third-party Dependency Versions Modified Since January 2021} & \np{171} & \np{150} \\ 
\bottomrule
\end{tabular}


\label{tab:descriptive}
\end{table*}

\subsection{Besu}

Besu is  the leading Java implementation for Ethereum \textsc{Eth1}.
Besu is led by the Hyperledger Foundation, a non-profit organization for open-source enterprise blockchain tools, which started in December 2015 as a spin-off of the Linux Foundation.
As of January 2022, there are at least \np{44} nodes of Besu running on the Ethereum Mainnet public network, according to Ethernodes.\footnote{\url{https://www.ethernodes.org}}
The source code of Besu is available on GitHub.\footnote{\url{https://github.com/hyperledger/besu}}
It is a reasonably sized and active project, containing a total of \np{268356} lines of code written in Java, contributed through \np{3125} commits. 
Besu is an active project, its codebase is developed and maintained by a total of \np{115} contributors with a unique GitHub account (of which \np{29} are listed as official maintainers).
The contributors reported and closed \np{916} issues and merged a total of \np{739} pull-requests in 2021.
More than half of the contributors work at Hyperledger, according to their GitHub profiles. 

In \autoref{tab:descriptive}, we capture some key statistics about the software supply chain of dependencies for Besu v21.10.6, released on January 5th, 2022. The raw data and analysis scripts are available online.\footnote{\url{https://github.com/chains-project/ethereum-ssc}}
We collected those dependencies using the Gradle dependencies' resolution plugin. 
On the analyzed release, Besu is made of \np{41} Gradle modules.
These modules are \emph{internal dependencies}, since their development, maintenance and release lifecycle is under the direct control of the Besu developers.
In addition to these \np{41} modules, Besu relies on \np{355} \textit{unique third-party dependencies} provided by \np{165} distinct supplying organizations. This number represents the number of different third-party Java libraries in the dependency tree of Besu, without considering the different versions of a dependency. The supplier organizations are in charge of maintaining these artifacts and releasing new versions to external repositories, with no formal ties with Hyperledger and Besu for most of them. 

In the central part of \autoref{fig:besu_node}, we zoom into the dependency tree of one of the \np{41} modules of Besu: \texttt{besu-core}. 
The compilation of this module depends on \np{29} internal dependencies, shown on the far left of the figure, as well as on \np{51} third-party dependencies that are also necessary for compilation.
The third-party dependencies are colored according to the name of the supplier organization.
For example, the third-party dependencies in dark yellow are handled by supplier ``Netty.''
Overall, the supply chain of \texttt{besu-core} is made of libraries maintained by \np{16} distinct suppliers.
We note that:
1) Many suppliers are large organizations with high code quality standards (\eg Apache, Google, JetBrains), which are trusted and relied on by many clients; 
2) There are partner suppliers such as the \texttt{quorum-mainnet-launcher} library developed by ConsenSys. Although not being maintained by Hyperledger, the developers are  close to the professional network of Besu developers;
3) Some libraries in the supply chain of \texttt{besu-core} belong to personal GitHub accounts, such as \texttt{picocli} and \texttt{snappy-java}. They are maintained and released by a single developer and cannot arguably be trusted as much as dependencies from big tech organizations or partner suppliers~\cite{Gustavsson20}.  

\subsection{Teku}

Teku is the leading \textsc{Eth2} Java node built to meet enterprise requirements. 
For example, Teku provides enterprise features such as monitoring with Prometheus, REST APIs for managing \textsc{Eth2} node operations, and external key management to handle validator signing keys.
Teku is an open-source project under active development on GitHub.\footnote{\url{https://github.com/ConsenSys/teku}}
The  first commit to the Teku codebase was made on September 9th 2018.
Since then, the project has seen a rapid development pace, accounting for \np{3142} commits contributed by a total of \np{65} developers.

\autoref{tab:descriptive} shows the descriptive statistics for the software supply chain of  Teku v22.1.0, released on January 3rd, 2022. 
The project contains a total of \np{57} unique internal dependencies and relies on \np{293} unique third-party dependencies.
Like Besu, Teku ships a large body of code coming from third-party dependencies with each new release. As for Besu, they are provided by \np{146} distinct suppliers with different code quality and security standards.
For the Ethereum ecosystem, the security and reliability of Besu's and Teku's supply chain are equally important.
One crashing bug or successful attack on either of them would potentially be devastating.

\subsection{Supply Chain Evolution}

In the bottom part of \autoref{tab:descriptive}, we give novel insights about the evolution of the Java Ethereum software supply chains.
We built the dependency trees of both supply chains, from January 2021: Besu v20.10.4 (\href{https://github.com/hyperledger/besu/tree/120d0d47f52641087299a5b51d43b69a9304dece}{120d0d4}) and Teku v21.1.0 (\href{https://github.com/ConsenSys/teku/tree/dcfb0ebeab457f818ec8b9afadf2faf78c4bd1e0}{dcfb0eb}). We compare these trees with  the versions released one year later, in January 2022.
We collect the number of dependencies introduced and modified in the supply chain of Besu and Teku, as well as the number of additional suppliers that have appeared along the evolution of these supply chains. 
We found \np{127} unique dependencies in the dependency tree of Besu and \np{79} dependencies in the dependency tree of Teku that are present in 2022 and that were not in the tree of 2021. This represents a significant growth of both supply chains, indicating the need for regular monitoring and assessment of the projects' dependencies.

The growth also holds for the number of suppliers of dependencies. In one year, there have been \np{49} and \np{22} new suppliers in the supply chains of Besu and Teku, respectively.
This is clear evidence that a complex supply chain evolves fast. 
Consequently, an approach based on “allow” and “deny” lists of suppliers is not viable, as it would necessitate frequent updates of these lists and potential delays in their assessment.
The management of the software supplier risks must be supported by tools that regularly monitor, analyze and assess the supply chain in order to cope with this evolution.

\subsection{Supply Chain Diversity}

\begin{figure}[t]
\centering
\includegraphics[width=7.3cm]{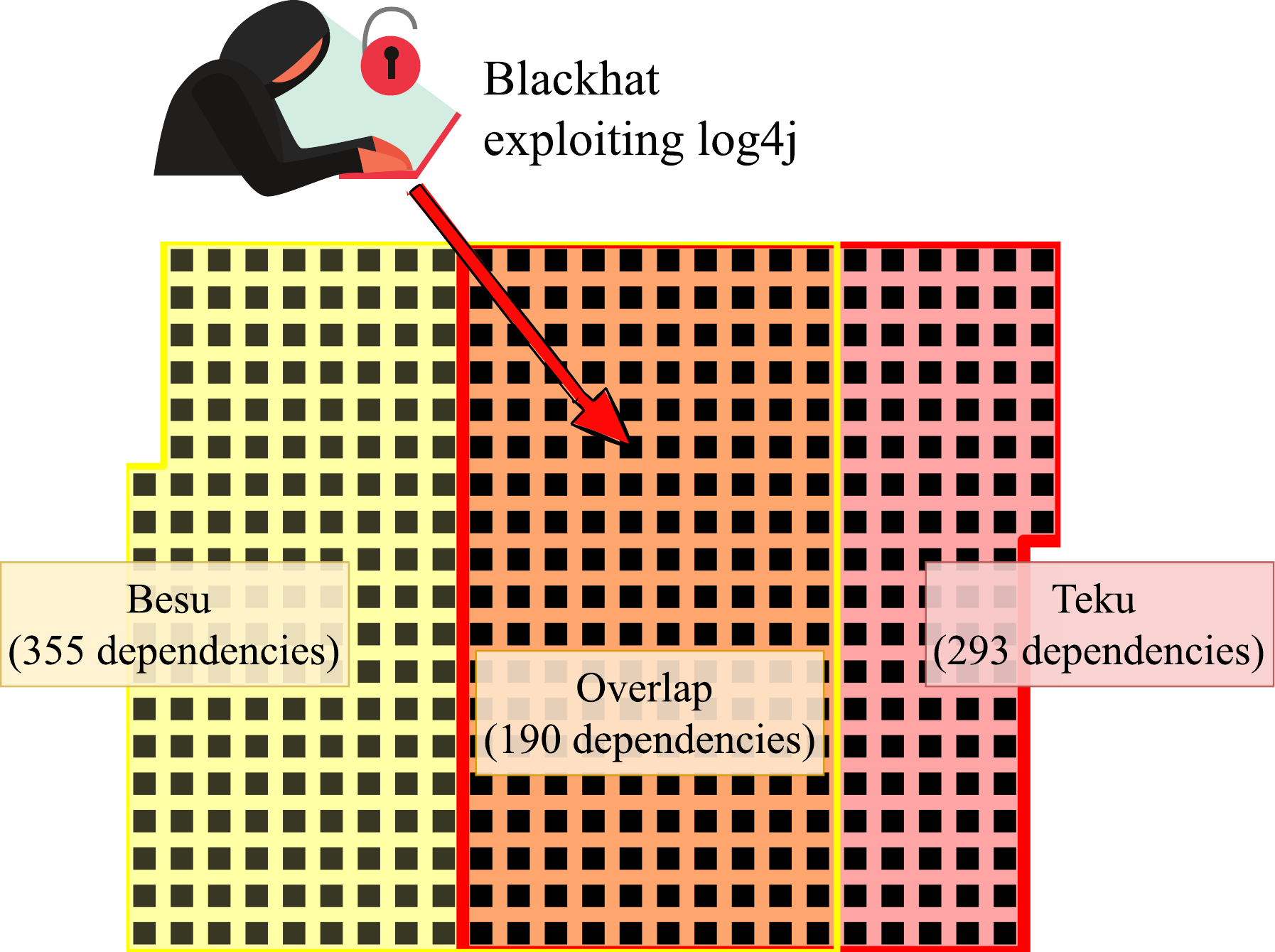}
\caption{Limited diversity among the software supply chains of Besu and Teku: the former includes 355 unique third-party dependencies and the latter is made of 293 dependencies, but 190 dependencies are shared by both supply chains. The shared dependencies represent 53.5\% and 64.8\% of the Besu and Teku supply chains.}

\label{fig:diversity}
\end{figure}

The Ethereum community values and explicitly promotes the development and the maintenance of a diversity of node implementations~\cite{Pooja2020}. Ethereum experts consider that node diversity is essential, in order for the network to be healthy and secure. 
Besu and Teku are two different node implementations, built by different development teams, following different development roadmaps. Meanwhile, their software dependencies represent a large body of their code base. We assess the diversity among these implementations by looking into the diversity among their dependencies and suppliers.
To do so, we extract the intersection of the dependencies of Besu and Teku. 
The supply chain of both nodes share a total of \np{190} third-party dependencies, which represents the \ShowPercentage{190}{355} and \ShowPercentage{190}{293} of the dependencies of Besu and Teku, respectively. 
This is illustrated in \autoref{fig:diversity}. 
Furthermore, we observe that \np{92} suppliers are common to both node implementations.
Even though Besu and Teku may look like entirely different node implementations, our results indicate that they actually carry out a large body of common code, a potential common failure point.
This suggests that the Ethereum community may work on supply chain diversity in addition to node diversity for further increasing resilience. 

Let us discuss the case of a dependency that is shared by both Besu and Teku: the Apache logging library  \texttt{log4j}.
In December 2021, a new CVE was published, documenting an exploit affecting all versions of \texttt{log4j} from version $2.0$ to $2.14.1$  (CVE-2021-44228)~\cite{Log4jCVE}. 
This caused a major disruption on the web, as \texttt{log4j} is a third-party dependency in thousands of software supply chains, including the ones of very critical services, such as Amazon and Microsoft Azure.
This vulnerability allows an attacker to perform arbitrary remote code execution on the running application, exploiting the vulnerable version of the \texttt{log4j} library.
Now, assume that an attacker had had the time to exploit this vulnerability in Ethereum Java nodes.
Since both nodes share the same dependency, it means that the scale of the repercussions would have been amplified.
The whole Ethereum ecosystem (both \textsc{Eth1} and \textsc{Eth2}) would have suffered from potential chain splits, violations of the consensus protocol,  and in the worst case, loss of funds and bored apes. 
If the two implementations had relied on diverse suppliers of logging facilities, the common failure risk would have been reduced. For example,  \texttt{Logback} or \texttt{Tinylog} are trustworthy alternatives to  \texttt{log4j}. 
Migrating from \texttt{log4j} to \texttt{Logback} in Besu requires minimal engineering effort: less than $10$ files need to be modified, thanks to modern Java logging architectures.
This diversification would benefit Besu nodes by providing different logging implementations from different suppliers, decreasing the chances of vulnerabilities with a blast effect.
We believe that the systematic assessment and enforcement of diversity in software supply chains is an important and promising research avenue.

\begin{table*}[]
\centering
\caption{Overview of risk metrics in the software supply chain of Besu v21.10.6 (\href{https://github.com/hyperledger/besu/tree/ef7984b}{ef7984b}) and Teku v22.1.0 (\href{https://github.com/ConsenSys/teku/tree/5b85ef1}{5b85ef1}). The data was obtained in January 15th, 2022.}
\begin{tabular}{@{}lcc@{}}
\toprule
                                 & \textsc{Besu (Eth1)}  & \textsc{Teku (Eth2)} \\ \midrule

\textsc{Outdated Third-Party Dependency Versions (Dependabot)} & \np{3}            & \np{1}                \\
\textsc{Outdated Third-Party Dependency Versions (Renovate)} & \np{49}            & \np{19}           \\
\midrule
\textsc{Vulnerable Third-Party Dependency Versions (OWASP)}     & \np{11}             & \np{2}      \\  
\textsc{Vulnerable Third-Party Dependency Versions (WhiteSource)}     & \np{15}             & \np{17}            \\ 


\bottomrule
\end{tabular}

\label{tab:actionable}
\end{table*}


\section{SUPPLY CHAIN REMEDIATIONS}

The two enterprise Java Ethereum nodes, Besu and Teku, depend on hundreds of third-party dependencies.
Today, there exist tools that can automatically enforce dependency management policies. Those policies include license checking, supplier approval, update frequency and security.
In this article, we focus on the solutions for the  latter two: identify outdated dependencies and replace vulnerable dependencies.

\subsection{Remediation of Outdated Dependencies}

Third-party libraries constantly  evolve to fix defects, patch vulnerabilities, and add features.
To take full advantage of third-party code, it is considered best practice to keep the dependencies up-to-date.
However, it is hard to stay up-to-date when the supply chain of a project includes a large number of dependencies, each of them having different lifecycles and release schedules.
In a large dependency tree, it is not uncommon that there is one new version of some dependency in the tree released \emph{per day}.

Keeping dependency up-to-date first means being aware of new versions (\eg due to a new release announcement or a security advisory). 
Once outdated dependencies are identified, the developers ensure that the update does not introduce breaking changes. Finally, they commit a change to bump the dependency version.
To our knowledge, the developers of Besu and Teku currently perform this monitoring and update procedure manually.
For instance, \autoref{lst:listing_besu_commons-codec} shows an example of a manual dependency update where a Besu developer updated the dependency \texttt{commons-codec} from version 1.13 to 1.15. 

\begin{lstlisting}[style=Java, language=diff, float, numbers=none, frame=trBL, belowskip=0\baselineskip, aboveskip=1\baselineskip, basicstyle=\footnotesize, caption={Example of a commit diff from a manual pull request (\href{https://github.com/hyperledger/besu/pull/3235}{PR \#3235}) made by a developer to update the dependency \texttt{commons-codec} in Besu.}, 
label={lst:listing_besu_commons-codec}]
@@ -49,7 +49,7 @@ dependencyManagement {
-   dependency `commons-codec:commons-codec:1.13'
+    dependency `commons-codec:commons-codec:1.15'
\end{lstlisting}

However, there exist software bots that automatically scan dependency trees and perform library updates, \eg Dependabot, Renovate, and Jared. 
To our knowledge, none of them are enabled in Besu and Teku. 
To assess their relevance in the context of these nodes, we performed a pilot experiment as follows.
We forked their GitHub repositories and configured Dependabot and Renovate to identify and remediate outdated dependencies.

\autoref{tab:actionable} shows the number of outdated third-party dependencies detected and reported by both dependency bots on January 15th, 2022.
Dependabot reports \np{3} and \np{1} outdated dependencies in Besu and Teku, respectively, whereas Renovate reports \np{49} and \np{19} outdated dependencies.
Renovate identifies many more updates because 1) it supports various package managers, 
and 2) it suggests updating infrastructure dependencies (in addition to application dependencies ).
Overall, both bots reveal several outdated dependencies that need to be acted upon.
This confirms that using these state-of-the-art supply chain tools would allow Besu and Teku developers to be more up-to-date and more diligent in handling their dependencies. From an economic perspective, it would avoid the engineering burden of manually checking new releases of their dependencies.

\subsection{Remediation of Vulnerable Dependencies}

Blackhat actors perform supply chain attacks~\cite{Pashchenko2018}. They purposefully compromise one dependency to achieve malicious goals, such as theft or denial of service.
Put it simply, vulnerable dependencies may cause a range of problems for Ethereum nodes related to their confidentiality, integrity, or availability.
Indeed, in December 2021, the Teku development team urgently mobilized their engineers after an important vulnerability disclosure: that of \texttt{log4j}~\cite{Log4jCVE} already mentioned above.
\autoref{lst:listing_teku_log4j} shows the commit made to fix this critical vulnerability.
To our knowledge, no successful attacks have been performed on Besu or Teku thanks to this timely commit.

As with outdated dependencies, technology exists to remediate vulnerable dependencies swiftly, such as OWASP Dependency Checker, Synk, or WhiteSource.
The identification of vulnerable dependencies in a software supply chain relies on scanning curated vulnerability databases, such as the National Vulnerability Database (NVD), and mapping vulnerability identifiers to versions in package repositories.
To our knowledge, the Teku team runs,as crontab job, a vulnerability identification tool called Trivly, but without automated remediation.

\begin{lstlisting}[style=Java, language=diff, float, numbers=none, belowskip=0\baselineskip, aboveskip=1\baselineskip, basicstyle=\footnotesize, frame=trBL, caption={Commit diff (\href{https://github.com/ConsenSys/teku/commit/a52f376b7020b8554f3b9e3e7f54c4158f9bfa6c}{a52f376}) showing a critical security update of the dependency \texttt{log4j} made to prevent a potential remote code exploit in Teku.}, 
label={lst:listing_teku_log4j}]
@@ -96,7 +96,7 @@ dependencyManagement {
-   dependencySet(group: `org.apache.logging.log4j', version: '2.13.3') {
+   dependencySet(group: `org.apache.logging.log4j', version: '2.15.0') {
      entry `log4j-api'
      entry `log4j-core'
      entry `log4j-slf4j-impl'
\end{lstlisting}


We searched for vulnerable dependencies in Besu and Teku with two state-of-the-art tools, considered as among the best tools in this domain: the OWASP Dependency Checker and WhiteSource. 
\autoref{tab:actionable} shows the results of the analysis, performed on January 15th, 2022.
OWASP detects \np{11} and \np{2} vulnerable dependencies in Besu and Teku, respectively, whereas WhiteSource detects \np{15} and \np{17} vulnerable dependencies. 
These results show that each tool focuses on different aspects, thus there is currently no silver bullet to identify dependency vulnerabilities. 
Interestingly, OWASP and WhiteSource both report the dependency \texttt{netty-transport} as affected by several vulnerabilities, which can be considered as a severe issue.
Also, we note that some vulnerable dependencies exists in both Besu and Teku, which is further evidence of the need for supply chain diversity discussed previously.

Neither Besu nor Teku use OWASP Dependency Checker or WhiteSource on a regular basis, yet.
Indeed, in January 2022, an active developer of Besu opened a pull-request to add the OWASP dependency checker in the build pipeline of the project (see \href{https://github.com/hyperledger/besu/pull/3288}{PR \#3288}, not merged at the time of writing).
Installing the OWASP Dependency Checker in the continuous integration pipeline of Besu would allow analyzing its dependency tree every time the node is built.
This way, developers are notified early in case of potential security issues related to third-party dependencies.
At the moment of writing this article, such an initiative has not been taken for Teku.
We believe that both Besu and Teku will eventually embed vulnerable dependency checking in their pipeline, this is inevitable for any major software project with a high stake.

\section{CONCLUSION} 

In this article, we deep-dived into the software supply chains of Besu and Teku. These two open-source projects are the major enterprise Java Ethereum nodes, which are in charge of financial and artistic transactions worth billions of dollars.
Our analysis reveals that both Ethereum node implementations are large software projects that depend on hundreds of libraries provided by a variety of supplier organizations.

Our work contributes to the state-of-the-art of software supply chains, with unique insights about complex networks of dependencies. We outlined the important growth of the software supply chains, as well as the increase of the number of suppliers on which Besu and Teku rely. We showed where the state-of-the-art lies with respect to remediation tools for hardening the software supply chain.     

While the Ethereum community stresses the importance of maintaining and incentivizing a diversity of node implementations, we have shown that the supply chains of Besu and Teku share a majority of their third-party dependencies. This is a serious limitation for the software diversity in the Ethereum ecosystem. Also, we have shown that dependency management for Besu and Teku can be improved with automated remediation. The significance of our findings  suggests that a similar analysis would be worthwhile for other Ethereum node implementations, such as Geth written in Go. 
Finally, we sincerely believe that our insights on the engineering of software supply chains hold for any blockchain that matters.

\section{ACKNOWLEDGMENTS}
\noindent This work has been partially supported by the Wallenberg Autonomous Systems and Software Program (WASP) funded by the Knut and Alice Wallenberg Foundation, as well as by the TrustFull and the Chains projects funded by the Swedish Foundation for Strategic Research.

\bibliographystyle{IEEEtran}
\bibliography{biblio.bib}
\balance
\newpage

\section{ABOUT THE AUTHORS}

\noindent\textbf{C\'esar Soto-Valero} is a PhD student in Software Technology at KTH Royal Institute of Technology in Stockholm, Sweden. His research work focuses on leveraging static and dynamic program analysis techniques to mitigate software bloat. Read about C\'esar on his personal website \href{https://www.cesarsotovalero.net}{https://cesarsotovalero.net}.\\

\noindent\textbf{Martin Monperrus} is Professor of Software Technology at KTH Royal Institute of Technology in Stockholm, Sweden.
He researches in software engineering, with a focus on software reliability.
He disseminates his research in industry through open-source and participation to industrial ventures.
He counsels the banking and financial sector with regard to distributed ledgers and digital assets.
Read about Martin at 
\href{https://www.monperrus.net/martin}{https://monperrus.net/martin}.\\

\noindent\textbf{Benoit Baudry}  is a Professor of Software Technology at KTH Royal Institute of Technology in Stockholm, Sweden.
He received his PhD in 2003 from the University of Rennes and was a research scientist at INRIA from 2004 to 2017.
His research in software engineering focuses on software testing and automatic diversification.
He also minted digital art on the Ethereum blockchain.
Read about Benoit on his personal website  \href{https://softwarediversity.eu}{https://softwarediversity.eu}.
\end{document}